\documentclass[twocolumn,aps,prl] {revtex4} 
\usepackage{graphicx}
\begin{document}
\pagestyle{empty} 
\title{Influence of frozen capillary waves on contact mechanics}
\author{B.N.J. Persson}
\affiliation{Oak Ridge National Lab, Condensed Matter Science Division,
Oak Ridge, Tennessee 37831, USA}
\affiliation{Kavli Institute for Theoretical Physics,
University of California Santa Barbara, CA 93106-4030, USA}
\affiliation{IFF, FZ-J\"ulich, 52425 J\"ulich, Germany}

\begin{abstract}
Free surfaces of liquids exhibit thermally excited 
(capillary) surface waves. 
We show that the surface roughness which results from capillary waves 
when a glassy material
is cooled below the glass transition temperature
can have a large influence on the contact mechanics between the solids. 
The theory suggest a new explanation for 
puzzling experimental results [L. Bureau, T. Baumberger and C. Caroli,
arXiv:cond-mat/0510232] about the dependence
of the frictional shear stress on the load for contact between a glassy polymer 
lens and flat substrates. It also lend support for a recently
developed contact mechanics theory. 
\end{abstract}
\maketitle


\vskip 0.5cm

Many technological applications require surfaces of solids to be extremely
smooth. For example, window glass has to be so smooth that no (or negligible)
diffusive scattering of the light occur (a glass surface with strong 
roughness on the length scale
of the light wavelength will appear white and non-transparent because
of diffusive light scattering). For glassy materials, e.g., 
silicate glasses, or glassy polymers, e.g., Plexiglas, 
extremely flat surfaces can be prepared by
cooling the liquid 
from well above the glass transition temperature $T_{\rm g}$, since in the liquid state 
the surface tension tends to eliminate (or reduce)
short-wavelength roughness. Thus,
float glass (e.g., window glass) is produced by letting melted glass flow into
a bath of molten tin in a continuous ribbon. The glass
and the tin do not mix and the contact surface between these
two materials is (nearly) perfectly flat. Cooling of the melt below the glass transition
temperature produces a solid glass with extremely flat surfaces.
Similarly, spherical particles with 
very smooth surfaces can be prepared by cooling liquid drops of glassy materials 
below the glass
transition temperature. Sometimes glassy objects are ``fire polished'' 
by exposing them to a flame
or an intense laser beam\cite{Lag} which melt a 
thin surface layer of the material, which 
will flow and form a very smooth surface in order to minimize
the surface free energy. In this way, small-wavelength roughness is
reduced while the overall (macroscopic) shape of the solid object is unchanged.  

However, surfaces of glassy materials prepared by cooling the liquid 
from a temperature above $T_{\rm g}$ cannot be perfectly
(molecularly) smooth, but exhibit a fundamental minimum surface roughness,
with a maximum height fluctuation  amplitude of typically $10 \ {\rm nm}$, 
which cannot be eliminated by any changes in the cooling procedure.
The reason is that at the surface of a liquid, fluctuations of vertical displacement are caused by
thermally excited capillary waves (ripplons)\cite{Jack,Die}. At the surface of very viscous 
supercooled liquids near the glass transition temperature these fluctuations
become very slow and are finally frozen in at the glass transition\cite{Jack}. 

The roughness derived from the frozen capillary waves is unimportant in many practical
applications, e.g., in most optical applications as is vividly evident for glass windows. 
However, in other applications they may
be of profound importance. In this letter I show that they are of crucial importance in
contact mechanics. For elastic hard solids such as silica glass, already a
roughness of order $1 \ {\rm nm}$ is enough to eliminate the adhesion, and reduce the contact area 
between two
such surfaces to just a small fraction of the nominal 
contact area even for high squeezing pressures\cite{P3}.
In this letter I will show that this is the case even for elastically much more compliant solids such
as Plexiglas (PMMA). The results presented below suggest a new explanation for the 
puzzling experimental result of Bureau et al\cite{Bureau} for PMMA, and
in addition lends support for a recently developed contact mechanics theory\cite{P8}.    
Bureau et al observed that the nominal shear stress $\sigma$ when a PMMA lens was slid on a 
flat substrate, depends strongly on the nominal pressure (or perpendicular stress) $p_{\rm nom}$.
They assumed that perfect (molecular) contact occurred in the nominal contact area $A_0$, and interpreted the
experimental results as a strong dependence of the frictional shear stress on the local pressure.
However, the analysis below shows that the experimental results can be understood assuming 
that the  area of real contact $A$ 
is smaller than the nominal contact area $A_0$, and assuming a {\it constant} (pressure independent)
shear stress $\sigma_{\rm f}$ in the area of real contact in such away that 
the friction force $F_{\rm f} = \sigma_{\rm f} A= \sigma A_0$ and the normal load $F_{\rm N} = pA=p_{\rm nom} A_0$,
where $p$ is the (average) pressure in the area of real contact.

The most important property of a rough surface is the surface roughness power spectrum
which is the Fourier transform of the height-height correlation function\cite{Nayak}:
$$C({\bf q})={1\over (2\pi )^2}\int d^2x \ \langle h({\bf x})h({\bf 0})\rangle e^{-i{\bf q}\cdot {\bf x}}$$
Here $z=h({\bf x})$ is the height of the surface at the point ${\bf x}=(x,y)$ above 
a flat reference plane chosen so that $\langle h({\bf x})\rangle = 0$. The angular
bracket $\langle ... \rangle$ stands for ensemble averaging.
The surface roughness power spectra due to capillary waves is of the form\cite{Die,Buzza,Sey}
$$C(q) = {1\over (2 \pi )^2} {k_BT\over \rho g +\gamma q^2 +\kappa q^4}\eqno(1)$$
where $\gamma$ is the surface tension, $\kappa$ the bending stiffness,
and $\rho$ the mass density of the glassy melt.
The smallest relevant wavevector $q = q_0 \approx 2\pi /L$, where $L$ is the diameter 
of the nominal contact area $A_0$. In the applications below $L\approx 0.1 \ {\rm mm}$
and (for PMMA) $\gamma \approx 0.04 \ {\rm J/m^2}$, giving $\gamma q^2 > 10^6 \ {\rm J/m^4} >> \rho g$.
Thus, the gravity term in (1) can be neglected. 
The mean of the square of the surface height fluctuation is given by
$$\langle h^2 \rangle = \int d^2q \ C(q) = 2\pi \int_{q_0}^{q_1} dq \ qC(q)$$
$$= {k_BT \over 2 \pi \gamma} {\rm ln} \left [ {q_1\over q_0} \left ({q_0^2+q_c^2 \over q_1^2+q_c^2}\right )^{1/2}\right ]\eqno(2)$$
where $q_c = (\gamma/\kappa)^{1/2}$ is a cross-over wavevector.
For the polymer surface which interests us below (PMMA), Eq. (2) gives the rms roughness $\approx 0.5 \ {\rm nm}$ which correspond
to a maximum roughness amplitude of about $3 \ {\rm nm}$. I will now show that even this small roughness has a
large influence on the contact mechanics.

\begin{figure}[htb]
   \includegraphics[width=0.4\textwidth]{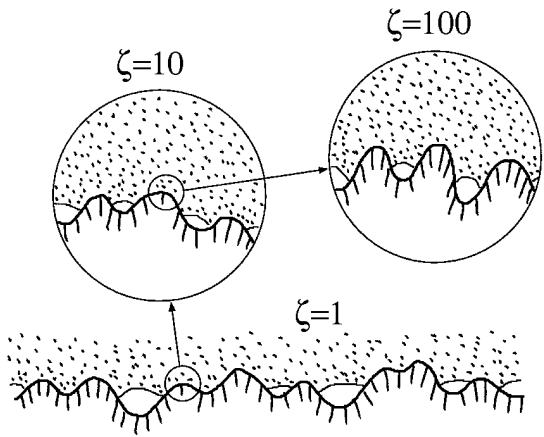}
\caption{
An elastic (e.g., rubber) block (dotted area) in adhesive contact with a hard
rough substrate (dashed area). The substrate has roughness on many different
length scales and the rubber makes partial contact with the substrate on all length scales.
When a contact area
is studied at low magnification ($\zeta=1$)
it appears as if complete contact occurs in the macro asperity contact regions,
but when the magnification is increased it is observed that in reality only partial
contact occurs.
}
\label{1x}
\end{figure}

Recently, a contact mechanics theory has been developed which is valid not only
when the area of real contact is small compared to the nominal contact area, but which is
particularly accurate when the squeezing force is so high that nearly complete contact occurs
within the nominal contact area\cite{P8,P1,P66}. All other contact mechanics theories\cite{Jon,GreenW,Bush} 
were developed for the case where the area of real contact is much smaller than the
nominal contact area. The theory developed in
Ref. \cite{P8,P1} can also be applied when the adhesional interaction is included.

Fig.~\ref{1x} shows the
contact between two solids at increasing magnification $\zeta$. At low magnification
($\zeta = 1$)
it looks as if complete contact occurs between the solids at many {\it macro asperity}
contact regions,
but when the magnification is increased smaller length scale roughness is detected,
and it is observed that only partial contact occurs at the asperities.
In fact, if there would be no short distance cut-off, the true contact area
would vanish. In reality, however,
a short distance cut-off will always exist since the shortest possible length is
an atomic distance. In many cases the local pressure at asperity contact regions
at high magnification will become so high
that the material yields plastically before reaching the atomic dimension.
In these cases
the size of the real contact area will be determined mainly by the yield stress
of the solid.

The stress distribution $P(\sigma, \zeta)$ at the interface 
when the interface is studied at the magnification 
$\zeta = L/\lambda$, where $L$ is the diameter of the nominal contact area 
between the solids and $\lambda$ the shortest surface roughness wavelength 
which can be detected at the resolution $\zeta$, satisfies the diffusion-like equation
$${\partial P \over \partial \zeta} = D(\zeta) {\partial^2 P \over \partial \sigma^2}\eqno(4)$$
where
$$D(\zeta)={\pi \over 4} \left ({E\over 1-\nu^2}\right )^2 q_0 q^3 C(q)\eqno(5)$$
where $E$ is Young modulus and $\nu$ the Poisson ratio, and where
$q_0=2\pi /L$ and $q=\zeta q_0$. The stress distribution function $P(\sigma , \zeta)$ must vanish for $\sigma < - \sigma_{\rm a}$,
where $\sigma_{\rm a} (\zeta)$ is the highest tensile stress possible at the interface when the system is
studied at the magnification $\zeta$. 
Substituting () in () gives
$$D(\zeta)={k_BT \over 16\pi\gamma} \left ({Eq_0\over 1-\nu^2}\right )^2 
{\zeta \over 1+\zeta^2 (q_0/q_c)^2} \eqno(6)$$
If one assume, as is expected from the theory of cracks, 
that the (tensile) stress diverge as $\sim r^{-1/2}$ with the
distance $r$ from the boundary line between a contact island and the surrounding non-contact region,
then one can easily prove 
the boundary condition 
$P(-\sigma_{\rm a}(\zeta),\zeta)=0$. When the adhesion is neglected $\sigma_{\rm a} = 0$ and 
$P(0,\zeta)=0$. 
The (normalized) area of (apparent) contact when the system is studied at the resolution $\zeta$ is given by
$${A(\zeta)\over \ A_0} = \int_{-\sigma_{\rm a}(\zeta)}^\infty d \sigma \ P(\sigma, \zeta)\eqno(7)$$
where $A_0$ is the nominal contact area.
The area of real contact is obtained from (7) at the highest (atomic) resolution,
corresponding to the magnification $\zeta_1=L/a$,
where $a$ is an atomic distance.

The detachment stress $\sigma_{\rm a}$ can be determined using the theory of cracks
which gives 
$$\sigma_{\rm a} (\zeta) \approx \left ({\gamma_{\rm eff} (\zeta) E 
q \over 2(1-\nu^2)}\right )^{1/2}\eqno(8)$$ 
Here $\gamma_{\rm eff} (\zeta)$ is the (effective) interfacial binding energy when the system is studied
at the magnification $\zeta$. At the highest magnification (corresponding to atomic 
resolution) $\gamma_{\rm eff}= \Delta \gamma$, where $\Delta \gamma = \gamma_1+\gamma_2-\gamma_{12}$ 
is the change in the energy (per unit area) when two {\it flat} 
surfaces of the solids are brought into adhesive contact. 
The effective interfacial energy at the magnification $\zeta$ is determined by the equation
$$A(\zeta) \gamma_{\rm eff}(\zeta ) =A(\zeta_1)\Delta \gamma - U_{\rm el}(\zeta)\eqno(9)$$ 
where $U_{\rm el}(\zeta)$ is the elastic energy stored at the interface due to the elastic 
deformation of the solid on length scales shorter than $\lambda = L/\zeta$, necessary in order to bring 
the solids into adhesive contact.
An explicit expression for $\gamma_{\rm eff}$ is given in Ref. \cite{P1}.

In the experiment by Bureau et al.\cite{Bureau}, the PMMA lens were prepared by cooling a liquid drop
of PMMA from $250 \ ^\circ {\rm C}$ to room temperature. In the liquid state the surface fluctuations of
vertical displacement are caused by thermally excited capillary waves (ripplons). When the liquid is near the
glass transition temperature $T_{\rm g}$ 
(about $100 \ ^\circ{\rm C}$ for PMMA), these fluctuations 
become very slow and are finally frozen in at the glass
transition. Thus, the temperature $T$ in (6) is 
not the temperature where the experiment was performed (room temperature), but rather
the glass transition temperature $T_{\rm g}$. 

\begin{figure}
  \includegraphics[width=0.45\textwidth]{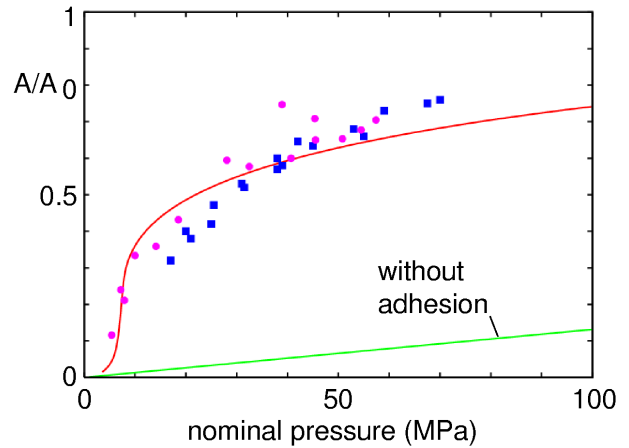}
  \caption{ \label{Baum}
The calculated relative contact area as a function of
the (nominal) pressure $p_{\rm nom}$ including adhesion (upper solid line)
and without adhesion. 
Circles: the normalized, 
(nominal) shear stress $\sigma /\sigma_{
\rm f}$ for
PMMA sliding on TMS as a function of the (nominal) pressure.
The nominal shear stress $\sigma$ has been divided by $\sigma_{
\rm f} = 50 \ {\rm MPa}$,
which is the (measured) shear stress in the area of real contact (see text).
Squares: the normalized, 
(nominal) shear stress $\sigma /\sigma_{
\rm f}$ (with $\sigma_{\rm f} = 5 \ {\rm MPa}$) for
PMMA sliding on OTS as a function of the (nominal) pressure.
}
\end{figure}

The (upper) solid line in Fig. \ref{Baum} shows the calculated relative contact area as a function of
the (nominal)  pressure $p_{\rm nom}=F_{\rm N}/A_0$ (where $F_{\rm N}$ is the squeezing force). 
In the calculation we have used the equations above 
with the measured elastic (Young) modulus $E=2.9 \ {\rm GPa}$
and surface tension $\gamma = 0.04 \ {\rm J/m^2}$. 
We have also used the glass transition temperature $T_{\rm g} \approx 370 \ {\rm K}$,
the short-distance cut-off wavevector $q_1= 7\times 10^9 \ {\rm m}^{-1}$
and $q_c =1.7\times 10^9 \ {\rm m}^{-1}$. 
The cross-over wavevector $q_c$ 
(or the bending stiffness $\kappa$) has not been measured for
PMMA, but has been measured for other systems. Thus, for alkanes at $T\approx 100 \ ^\circ{\rm C}$
for C20 and C36, $q_c \approx 4.4\times 10^9$ and $\approx 2.7 \times 10^9 \ {\rm m}^{-1}$,
respectively\cite{Ocko}. There are also some studies of polymers
\cite{Bollinne,Jonas} (using the atomic force microscope)
giving $q_c \approx 1\times 10^8 \ {\rm m}^{-1}$. 
The interfacial binding energy $\Delta \gamma$ can be estimated 
using\cite{Is} $\Delta \gamma \approx 2 (\gamma_1 \gamma_2)^{1/2}
\approx 0.07 \ {\rm J/m^2}$, where $\gamma_1=\gamma \approx 0.04 \ {\rm J/m^2}$ is the surface
energy of PMMA and $\gamma_2 \approx 0.02 \ {\rm J/m^2}$ the surface energy of 
the (passivated) substrate. In the calculations we used $\Delta \gamma = 0.06 \ {\rm J/m^2}$.

If one assumes that, in the
relevant pressure range (see below), 
the frictional shear stress $\sigma_{
\rm f}$ between PMMA and the substrate is essentially {\it independent} of the normal stress $p$ 
in the asperity contact regions, 
then the calculated curve in Fig. \ref{Baum} 
is also the ratio between the nominal (or apparent) shear stress $\sigma$ and the
true stress $\sigma_{
\rm f}$; $A/A_0 = \sigma /\sigma_{\rm f}$ 
(note: the friction force $F_{\rm f} = \sigma_{\rm f} A = \sigma A_0$). 

The circles and squares in Fig. \ref{Baum} show the measured data of
Bureau et al.\cite{Bureau} for $\sigma /\sigma_{
\rm f}$. They performed
experiments where a PMMA lens, prepared by cooling a liquid drop of PMMA,
was slid on silicon wafers (which are nearly atomically smooth)
covered by a grafted silane layer. Two different types of 
alkylsilanes were employed for surface modification, namely a trimethylsilane (TMS)
and octadecyltrichlorosilane (OTS). The
circles in Fig. \ref{Baum} are the 
shear stress,
divided by $\sigma_{
\rm f} = 50 \ {\rm MPa}$, for
PMMA sliding on TMS as a function of the (nominal) pressure $p_{\rm nom}$.
The shear stress  $\sigma_{
\rm f} = 50 \ {\rm MPa}$ was deduced from multi-contact 
experiments\cite{Bureau1} 
(where the local pressure in the asperity regions is so high as to give rise to local
plastic deformation) using
$\sigma_{
\rm f} \approx \sigma H/p_{\rm nom}$, where $H \approx 300 \ {\rm MPa}$ is the hardness 
(yield stress) of PMMA
as determined from indentation experiments. 
This equation follows from $\sigma A_0 = \sigma_{\rm f} A$ and $A H = A_0 p_{\rm nom}$.
The squares in Fig. \ref{Baum} show 
$\sigma /\sigma_{
\rm f}$ for
PMMA sliding on OTS as a function of the (nominal) pressure $p_{\rm nom}$.
For this system no multicontact experiments were performed, and we have 
divided the apparent shear stress by  $\sigma_{
\rm f}=5 \ {\rm MPa}$, chosen so as to
obtain the best agreement between the theory and experiment. Hence, in this case
$\sigma_{
\rm f}$ represent a theoretically predicted shear stress for the multicontact
case; we suggest that multicontact experiments for the system PMMA/OTS 
are performed in order to test this prediction. 

The calculation presented above shows that the local pressure in the asperity
contact regions is of order $\sim 2 \ {\rm GPa}$, which is much higher than the 
macroscopic yield stress (about $300 \ {\rm MPa}$) of PMMA. 
However, nanoscale indentation experiments\cite{Japan} have shown
that on the length scale ($\sim 10 \ {\rm nm}$) and indentation depth scale ($\sim 1 \ {\rm nm}$)
which interest us here, the indentation hardness of PMMA is of order $\sim 10 \ {\rm GPa}$,
and no plastic yielding is expected to occur in our application.

The study above has assumed that the frictional shear stress at the
PMMA-substrate interface is independent of the local pressure $p$,
which varies in the range $p \approx 50-100 \ {\rm MPa}$. This result is expected
because the shear stress will in general exhibit a negligible pressure
dependence as long as $p$ is much smaller than the {\it adhesional}
pressure $p_{\rm ad}$. The adhesional pressure is defined as follows: In order for
local slip to occur at the interface, the interfacial molecules must pass over
an energetic barrier of magnitude $\delta \epsilon$. At the same time the spacing
between the surfaces must locally increase with a small amount (some fraction of an
atomic distance) $\delta h$. This correspond to a pressure work $p a^2 \delta h$,
where $a^2$ is the contact area between the molecule, 
or molecular segment, and the substrate.
Thus the total barrier is of order $\delta \epsilon + p a^2 \delta h =
a^2 \delta h (p_{\rm ad}+p)$ where $p_{\rm ad} = \delta \epsilon / (a^2 \delta h)$.
For weakly interacting systems one typically 
have\cite{last} $\delta \epsilon /a^2 \approx 1 \ {\rm meV}/{\rm \AA}^2$
and $\delta h \approx 0.01 \ {\rm \AA}$ giving $p_{\rm ad} \approx 10 \ {\rm GPa}$, 
which is at least one order of magnitude larger than the pressures in the present experiment. 
We note that the pressure $p_{\rm ad}$ is similar to the  
pressure $p$ which must be applied 
before the viscosity of liquid hydrocarbon oils start to depend on the applied pressure.
The reason for the pressure dependence of the
viscosity of bulk liquids is similar to the problem above, involving the formation of
some local ``free volume'' (and the associated work against the applied pressure)
in order for the molecules to be able to rearrange during shear.

To summarize, 
I have shown that the surface roughness which result from capillary waves 
when a glassy material
is cooled below the glass transition temperature
can have a large influence on the contact mechanics between the solids,
even for relatively (elastically) soft solids such as PMMA. 
This fact suggest a new explanation for puzzling experimental results about the dependence
of the frictional shear stress on the load for contact between a PMMA 
lens and flat passivated silica substrates. It also lend support for a recently
developed contact mechanics theory. 

\vskip 0.3cm
{\bf Acknowledgement}
I thank T. Baumberger and L. Bureau for drawing my attention to their interesting experimental
data and for supplying me the data used in Fig. 1.
I thank T. Baumberger, L. Bureau, C. Caroli, R. Carpick, M. Miguel and M. M\"user 
for useful comments on the manuscript.
This work was performed during a three month stay at ORNL and a one month stay at the
KITP. This  
work is partially supported by 
Oak Ridge National Laboratory, managed 
by UT-Battelle, LLC, for the U.S. Department 
of Energy under Contract No. DE-AC05-00OR22725,
and by the 
National Science Foundation under Grant No. PHY99-07949.

\end{document}